\documentclass[aps,prb,twocolumn,showpacs,groupedaddress]{revtex4} 

\usepackage{graphicx}
\usepackage{dcolumn}
\usepackage{bm}
\usepackage{amsmath, amsfonts, amssymb}
\usepackage{amsxtra}
\usepackage{amsthm}
\usepackage[usenames]{color}

\usepackage{soul} 

\newcommand{\eqnref}[1]{Eq.~(\ref{#1})} 
\newcommand{\fref}[1]{Fig.~\ref{#1}}    

\newcommand{\beq}{\begin{equation}}
\newcommand{\eeq}{\end{equation}}
\newcommand{\ba}{\begin{array}{ccc}}
\newcommand{\ea}{\end{array}}

\def\bea{\begin{eqnarray}}
\def\eea{\end{eqnarray}}

\def \be{\begin{equation}}
\def \ee{\end{equation}}
\def \ba{\begin{array}}
\def \ea{\end{array}}
\def \bea{\begin{eqnarray}}
\def \eea{\end{eqnarray}}

\def \half{\frac{1}{2}}
\def \etal{{\it {et al}}}

\def \W{{\Omega}}

\def \a{{\alpha}}
\def \t{{\theta}}

\newcommand*{\diff}[3][]{\frac{d^{#1}{#2}}{d^{}{#3}{}^{#1}}}

\usepackage{setspace} 
\begin{document}

\title{Numerical evidence for strong randomness scaling at a superfluid-insulator transition of one dimensional bosons}

\author{Susanne Pielawa$^1$ and Ehud Altman$^{1,2}$}
\affiliation{
$^1$Department of Condensed Matter Physics, Weizmann Institute of Science, Rehovot 76100, Israel\\
$^2$Department of Physics, University of California, Berkeley, California 94720, USA
}


\date{\today}

\begin{abstract}
We present numerical evidence from Monte-Carlo simulations that the superfluid-insulator quantum phase transition of interacting bosons subject to strong disorder in one dimension is controlled by the strong-randomness critical point. 
At this critical point the distribution of superfluid stiffness over disorder realizations develops a power-law tail reflecting a universal distribution of weak links. The Luttinger parameter on the other hand does not take on a universal value at this critical point, in marked contrast to the known Berezinskii-Kosterlitz-Thouless-like superfluid-insulator transition in weakly disordered systems.  We develop 
a finite-size scaling procedure which allows us to directly compare the numerical results from systems of linear size up to 1024 sites with theoretical 
predictions obtained in 
[PRL {\bf{93}}, 150402 (2004)],
using a strong disorder renormalization group approach. 
The data shows good agreement with the scaling expected at the strong-randomness critical point.  
\end{abstract}
\maketitle

\section{Introduction}
Superfluids at non vanishing temperature in two dimensions, or at zero temperature in one dimension, cannot exhibit broken symmetry~\cite{PhysRev.158.383,Mermin1966}. Instead they may sustain at best algebraic off-diagonal correlations. 
Establishment of algebraic order at a critical temperature, or in one dimension, at a critical value of some Hamiltonian parameter, is ubiquitously understood in terms of the Berezinskii-Kosterlitz-Thouless transition (BKT)~\cite{0022-3719-6-7-010, Berezinskii}. The hallmark of a BKT transition is a universal jump of the exponent associated with the algebraic decay, or Luttinger parameter, at the critical point~\cite{PhysRevLett.39.1201,PhysRevB.16.1217}. Can the transition occur 
through an alternative mechanism, which would give rise to a different universality class?

Systems of bosons in a disordered one-dimensional potential provide a good arena to investigate this question. Since  both the effects of interaction and of quenched disorder are particularly strong in one dimension, the interplay between the two may well lead to new physics. Moreover, such systems naturally occur in a rather large variety of physical realizations that are currently under intense study. These include  ultra-cold atomic experiments, where Anderson localization of non-interacting bosons has been observed~\cite{Roati2008a,Billy2008}, disordered superconducting quantum wires \cite{Bezryadin2000}, and Josephson junction arrays~\cite{Haviland2000}.

In a seminal work, Giamarchi and Schulz~\cite{Giamarchi1987}  addressed the localization transition of interacting bosons by starting from a clean harmonic fluid (Luttinger liquid) description and added the disorder perturbatively. This analysis, as well as a recent two-loop calculation\cite{Ristivojevic2012a}, predict a BKT transition, but with a universal jump of the Luttinger parameter that is different than the clean case.

However, the natural regime of weakly interacting cold atomic Bose gases was shown to be very far from the weakly disordered regime: 
a weakly-interacting quasi-condensate in a disordered potential forms superfluid puddles which are coupled by strongly disordered Josephson couplings\cite{Vosk2012, PhysRevLett.98.170403, PhysRevB.80.104515}.  

Such a model has been argued to offer an alternative paradigm for the superfluid insulator transition, in which the strong disorder plays a central role~\cite{AKPR,AKPR2,Altman2009}. Analysis of the problem using strong disorder renormalization group (SDRG) found a transition that is controlled by rare weak links that can potentially cut the chain.  In the course of renormalization to longer scales, the effective distribution of weak links flows toward a universal structure, which determines the universal behavior at the critical point. The value of the Luttinger parameter at the transition is, however, non-universal and depends on where the phase boundary is crossed~\cite{AKPR,AKPR2,Altman2009,2013arXiv1307.7719} (i.e. on the strength of the bare disorder and interactions). 

The question if the proposed critical point is actually realized remains controvertial~\cite{Balabanyan2005,Pollet2013a}. This controversy is exacerbated by the fact that while the disorder grows under the SDRG flow it does not reach an infinite value and thus the method does not become asymptotically exact as  it does in other instances of random critical points~\cite{Fisher1994}. It is therefore hard to rule out the possibility that even when the bare disorder is strong the critical system eventually flows to the weak disorder Giamarchi-Schulz fixed point. 

A number of previous works have attempted to resolve this question numerically, but the lack of an appropriate finite-size scaling procedure did not allow to probe directly for the strong-disorder fixed point.  
Balabanyan \etal~\cite{Balabanyan2005} found relatively good agreement with the BKT transition using one type of disorder, but noted strong finite-size effects when a different disorder type was used. 
Hrahsheh and Vojta~\cite{Hrahsheh2012a} measured a non-universal Luttinger parameter at the transition that is larger than the value predicted by a BKT critical point, consistent with the prediction of the SDRG. On the other hand, Pollet \etal~\cite{Pollet2013a} proposed a fit of numerical data assuming a fully classical scaling of the stiffness with system size, unaffected by the quantum effects of the charging energy. They argued that the Luttinger parameter found in their numerical simulation and extrapolated to the thermodynamic limit using the aforementioned classical scaling does not exceed the value predicted by the standard Giamarchi-Schulz theory at the critical point.
Neither of these works probed directly for the universal predictions of the strong disorder theory. 

In this paper we characterize the critical properties of the transition using Quntum Monte Carlo simulations of a standard bosonic model,  comparing directly against universal predictions of the SDRG. The crucial new element in our analysis is the development of an appropriate finite size scaling procedure. We find good agreement with the scaling predicted by the SDRG at the random critical point. Even inside the superfluid phase, at strong disorder, the distribution of relevant quantities, such as the inverse superfluid stiffness
\footnote{the superfluid stiffness is the response of the system to a phase twist $\theta$ at the boundaries. It is defined by the second derivative of the minimal energy with respect to the phase difference  $\rho_s=\frac 1 L \frac{\partial^2 E}{\partial \theta^2}$.} 
is seen to develop broad tails which may  invalidate the weak disorder analysis of Giamarchi and Schulz~\cite{Giamarchi1987}. We thus conclude that beyond a certain disorder strength the transition changes character, and is controlled by the strong-randomness fixed point. In an appendix we address the issue of extrapolating a Luttinger parameter from finite size data. We show explicitly that the classical scaling formula used in Ref. \onlinecite{Pollet2013a} significantly underestimates the true value of the Luttinger parameter.

\begin{figure}[tb]
\begin{center}
\includegraphics[width=\columnwidth]{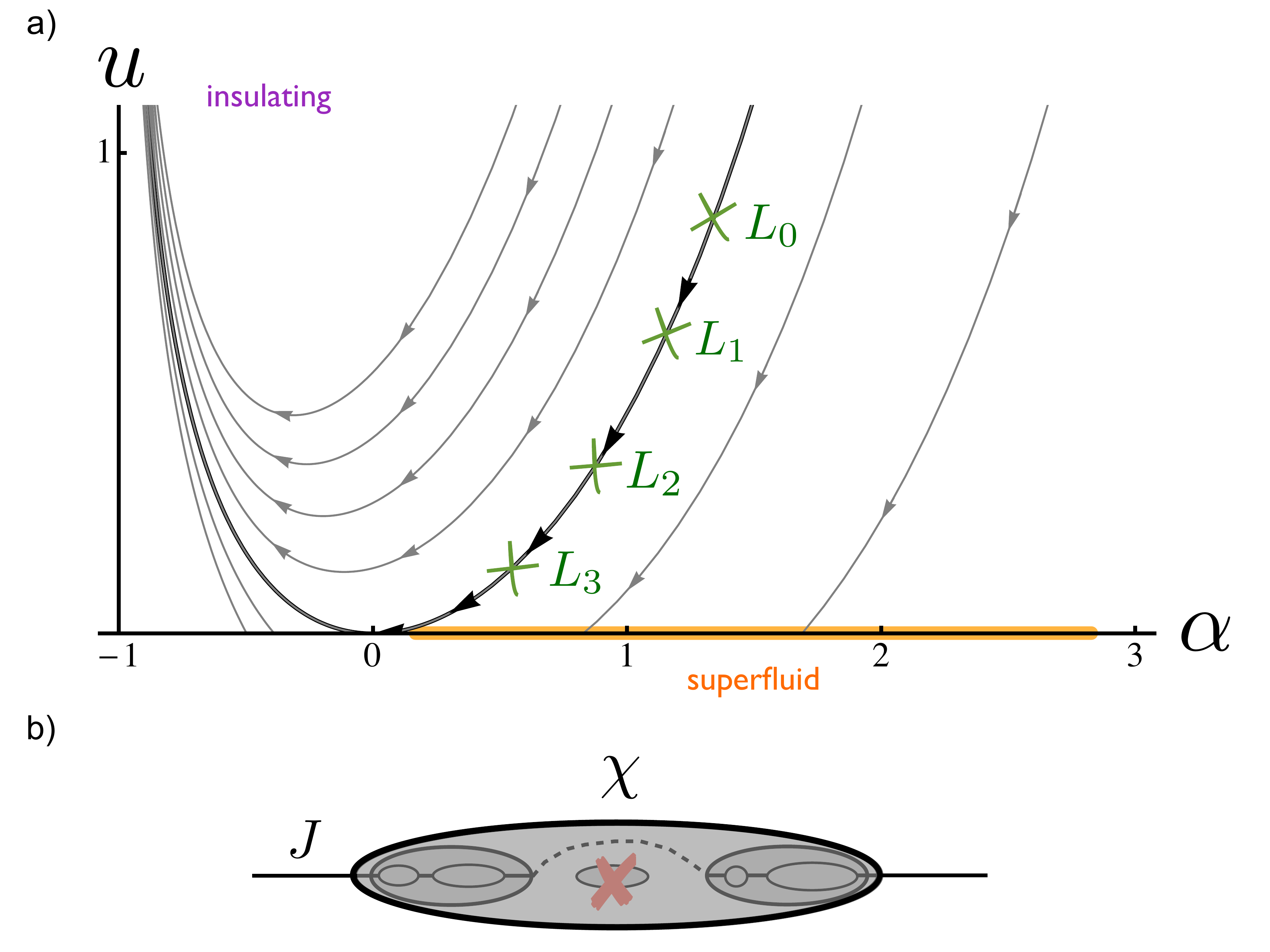}
\end{center}
\caption{
a) Solutions to the flow equations near the strong-disorder fixed point. 
The parameter $\alpha$ characterizes the renormalized probability distribution of weak links, $p(J)\propto J^\alpha$. 
This parameter $\alpha$ can be extracted from numerical or experimental data via the distribution of superfluid stiffness. 
The parameter $u$ describes the renormalized distribution of on-site repulsion and cannot be directly extracted from physical observables. 
%
The flow proceeds along the flow lines in the direction of the arrows, i.e. if a system starts out with bare parameters $\alpha_0$ and $u_0$ (point $L_0$ in the graph) at system size $L_1$ we will measure $\alpha_1$. At a larger system size $L_2$ we will measure $\alpha_2$, and so on. 
%
b) The flow stops when the entire chain has shrunk to a single superfluid cluster. Here the last couple of SDRG steps are shown schematically. The last cluster has an internal stiffness $\chi$ and is connected to itself by the last remaining link $J$. Both $\chi$ and $J$ are combined to obtain the total superfluid stiffness of the system, see \eqnref{eq:stiffness}.
}
\label{fig1}
\end{figure}

\section{From SDRG to measurable quantities} 
In order to establish the existence of a strong disorder superfluid-insulator critical point we need to identify observables, measurable in Monte Carlo simulations, that are predicted by SDRG to obey distinct universal behavior. 
The SDRG was carried out in Refs.~\onlinecite{AKPR,Altman2009} on the following Josephson array model
\be
H=\sum_i \half U_i n_i^2 -J_i\cos(\t_{i+1}-\t_i)-\mu_i n_i,
\label{JJA}
\ee
with $U_i$ random charging energies of grains, $J_i$ Josephson couplings and $\mu_i$ a random chemical potential giving rise to offset charges. For simplicity we will focus here on the case of a particle-hole symmetric model (i.e. $\mu_i=0$), however, the same analysis carried out on a model with generic disorder leads to essentially the same conclusions about the transition at strong disorder. 

\begin{figure*}[tb]
\begin{center}
\includegraphics[width=\columnwidth]{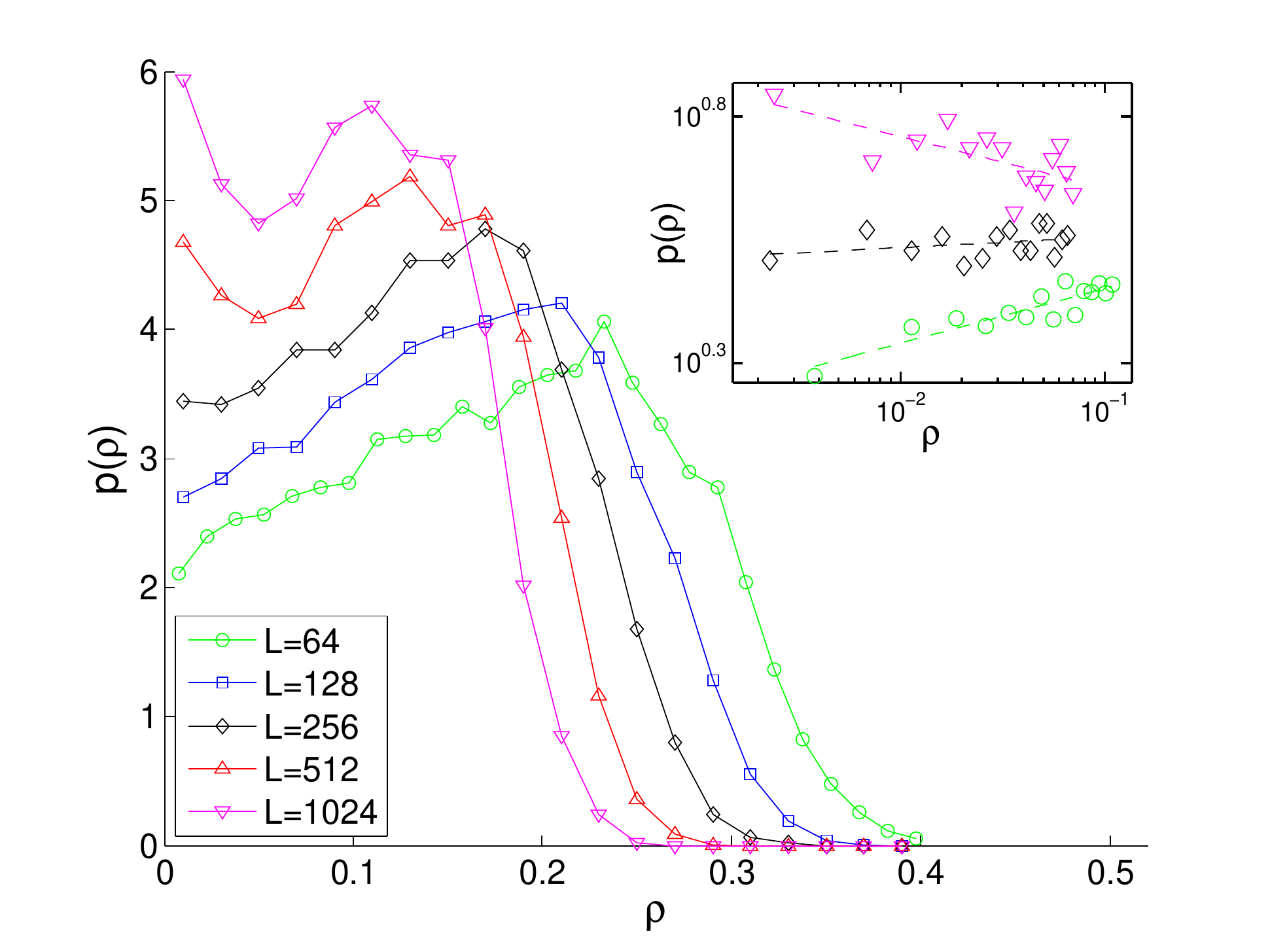}
\includegraphics[width=\columnwidth]{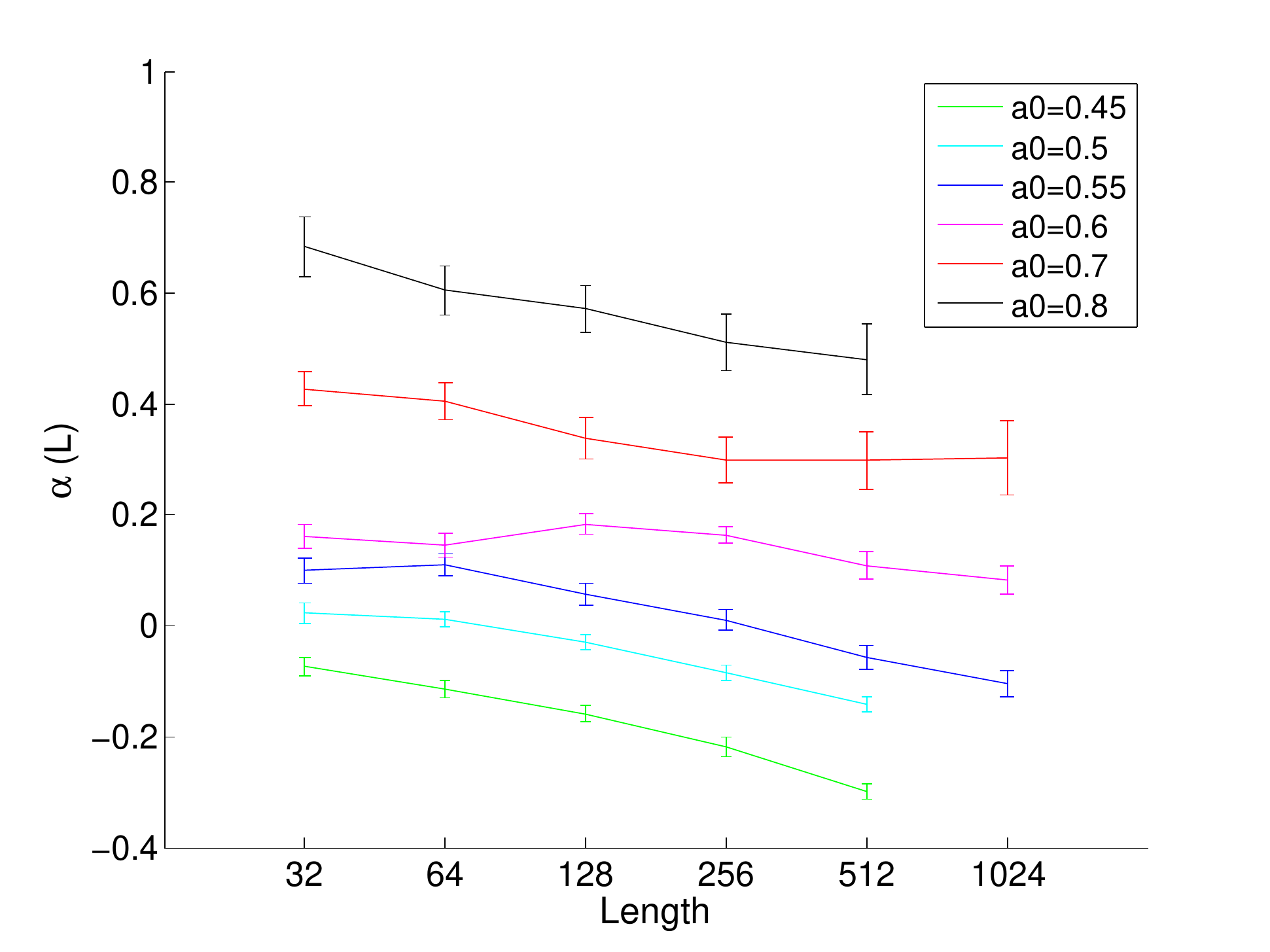}
\end{center}
\caption{
a) Distribution of superfluid stiffness for $\alpha_0=0.55$ for different system sizes. Near $\rho=0$ the distribution follows a power law, which changes as the systems size is increased.
The inset shows the near-zero-tail on a log log scale with finer binning to visualize the power-law. A maximum-likelyhood-estimator was used to extract the power $\alpha$, see text. 
b) Values of $\alpha$ extracted for different systems sizes $L$  and initial conditions $\alpha_0$, on a semi-log plot. 
This shows that the distribution of weak links flows as the system size is increased: more weak links are generated.
}
\label{fig2}
\end{figure*}

In the course of renormalization successive elimination of the largest couplings on the chain (\ref{JJA}), leads to a flow toward universal distributions of coupling constants
\begin{subequations}
\bea
p(U)&=& \frac{e^U}{\Omega} \left( \frac \Omega U\right)^2 e^{-u\Omega/U}\\
p(J)&=&\frac {\alpha+1}{\Omega}\left( \frac J \Omega \right)^\alpha
\eea
\end{subequations}
Hence in late stages of the RG the flow of full distribution functions is encapsulated in scaling equations for only two parameters $u$ and $\alpha$. 

The resulting flow is shown in \fref{fig1}. Below the separatrix the flow approaches the superfluid fixed line on the $\a$ axis. The transition to the insulator occurs when the renormalized value of $\a$ takes on a universal value $\a=0$. That is, when the renormalized distribution of Josephson couplings becomes uniform. For $\alpha<0$ weak links become dominant and effectively cut the chain.  

The essential difficulty in identifying this universal behavior in numerical simulations is that the object which becomes universal, i.e. the renormalized distribution of Josephson couplings, is a  theoretical construct and not directly measurable. On the other hand,  averages of physical quantities, such as the superfluid stiffness~$\rho$, compressibility~$\kappa$, or Luttinger parameter $K$, take on non-universal values at the strong disorder critical point~\cite{Altman2009}. Our first task will therefore be to relate the renormalized disorder distributions with a measurable quantity.
%
%
As we show below the right quantity to consider is the distribution of superfluid stiffnesses of an ensemble of realizations of the disordered chain. We argue that the tail of this distribution near $\rho=0$ follows a power law with the desired exponent $\alpha$.

This can be seen as follows. The superfluid stiffness of a chain of length $L$ is computed by proceeding with the SDRG until the entire chain has shrunk to a single superfluid cluster~\cite{Altman2009}. The last value of $J$ is then the coupling of this cluster to the leads, or to itself for periodic boundary conditions. 
The superfluid stiffness $\rho$ for a single disorder realization is calculated as 
\begin{equation}
\frac 1 {\rho}=\frac{1}{L}\left( \chi+ \frac{1}{J}\right),
\label{eq:stiffness}
\end{equation}
where $\chi=\sum_{i\in \rm{cluster}} {J_i}^{-1}$ keeps track of the stiffness of the superfluid cluster. Here $J_i$ are bonds that were joined to create the cluster~\cite{Altman2009}. 
Since $\chi$ is integrated over the entire flow it is not universal and depends on the initial disorder.
%
%
All energy scales appearing in $\chi$ are above the cutoff, while $J$ is still below the cutoff: it is the last link that has not yet been decimated. For samples that have a small $\rho$ the largest contribution must have come from this last link, J. Therefore, at that energy scale, the tail of the distribution of the superfluid stiffnesses  near $\rho\approx 0$ follows the same power law with exponent $\alpha$ as does $P(J)$. For larger values of $\rho$ the distribution becomes non-universal due to the contribution of $\chi$. 

\section{Numerical simulations}  
We shall perform numerical simulation of the directed loop model~\cite{Wallin1994c, Balabanyan2005}, which is presumed to have the same universal properties as the Josephson array (\ref{JJA}).
This model is given in terms of the lattice action
\begin{equation}
S=\half\sum_{\vec{n}} 
\left(
K_{\vec n, \hat x} \, J^2_{\vec n, \hat x}
+
K_{\vec n, \hat \tau} \, J^2_{\vec n, \hat \tau},
\label{loop}
\right).
\end{equation}
where the weights are assigned according to the covering of the lattice with directed loops of currents $J_{\vec n, \a}$ on the bonds. The integer vector $\vec n = (x, \tau)$ labels the sites of the two-dimensional square lattice, $\hat x$ and $\hat \tau$ are unit vectors in the two lattice directions. The allowed configurations for the currents fulfill a zero-divergence constraint $\sum_\alpha\left( J_{\vec n, -\alpha} + J_{\vec n, \alpha}\right)=0$, 
and by definition $J_{\vec n, -\alpha}=- J_{\vec n - \alpha, \alpha}$. 

As usual the extra dimension of the classical model ($\tau$ axis) stems from the imaginary time axis in the path-integral representation of the quantum partition function. Therefore the coupling constants must be uniform along that axis while they vary randomly along the $x$ axis. These couplings can be roughly related to those of the original Josephson array model through $K_\tau=U$ and $K_x=-2\ln(J/2)$~\cite{Balabanyan2005}.


We take bare coupling distributions that are not far from the expected fixed point distributions in order to have fast finite size convergence to the universal regime. For the disorder in the parameter $K_\tau \approx U $, we take a distribution with $u_0=0.2$: 
$p(K_\tau)=\left( {e^{u_0}}/{K^2_\tau} \right)e^{-u_0/K_\tau}$ for $0<K_\tau<1$.  
The bare-disorder distribution for the parameter $K_x$ 
is taken as $p(K_x)=2^{\alpha_0}(\alpha_0+1) e^{-(\alpha_0+1)K_x/2}$ for $K_x > 2 \ln 2$. By change of variable this is a power-law distribution for the Josephson coupling $J$ with a power $\a_0$ used here as the tuning parameter of the transition. 

We performed Monte Carlo simulations of the model~(\ref{loop}) using a classical worm algorithm~\cite{Prokof'ev2001}. For each disorder realization we computed the stiffness $\rho_x$ \footnote{$\rho_x$ is the stiffness of the one dimensional quantum model. The stiffness of the two dimensional model $K=\pi \sqrt{\rho_x\rho_\tau}$ defines the Luttinger parameter.} and constructed a full distribution from a set of 3000 to 20000 disorder realizations for each value of $\a_0$ and system length $L$~\footnote{Only a fraction of the disorder realizations contribute to the tail of the distribution which determines $\a(L)$. This fraction is smaller near the transition than in the insulator and even smaller deep in the superfluid because less weak links are being generated. Hence more disorder realizations are needed the more we enter into the superfluid.}. To obtain the renormalized value of $\alpha(L)$ we extracted the exponent of the power law at the tail of the distribution at small $\rho$ using a maximum-likelyhood estimator~\cite{doi:10.1137/070710111}.

The results of this calculation are presented in \fref{fig2}. 
First, \fref{fig2}(a) shows the stiffness distributions obtained for one value of the bare disorder $\a_0$ and varying system sizes. In this case it is clear that the power associated with the tail of the distribution flows from slightly positive value to a negative one, i.e. to a diverging probability density at $\rho=0$. \fref{fig2}(b) shows the change of the exponent $\a$ with $L$ for varying values of the tuning parameter $\a_0$.
The error bars for $\alpha$ shown on the plot stem from two independent sources: a) the QMC error due to limited simulation time; it is obtained by creating data sets of values for $\rho$ shifted within their QMC error bars, then recalculating $\alpha$.  b) An error due to limited number of disorder realizations; this error is estimated as described in Ref.~\onlinecite{doi:10.1137/070710111}. 

At this point we can already make a few interesting observations about the transition. 
First, note that the tail of the stiffness distribution ubiquitously flows with system size $L$, giving rise to a power-law tail with an exponent $\alpha$ that decreases monotonically with $L$. Within the SDRG this is understood as being due to renormalization of the Josephson coupling on eliminating sites with large charging energy. Hence, while the flow of average stiffness may appear to be approximately classical~\cite{Pollet2013a}, the flow of the full distribution and in particular its tail has an important quantum contribution. The second  observation is that even in the superfluid phase the superfluid stiffness retains a power-law distribution near $\rho=0$ with exponent $\alpha<1$. Consequently the variance of $1/\rho$ diverges in this regime. This has important implications because a finite variance of $1/\rho$ is required to obtain a dual description in terms of phase slip variables as a starting point of a weak disorder analysis ~\cite{Giamarchi1987}. Hence the numerically obtained stiffness distributions can by themselves explain the breakdown of the weak disorder theory. 

\section{Finite-size scaling}
To quantitatively assess the agreement between the numerical data and the universal behavior expected in the vicinity of the strong disorder critical point~\cite{AKPR,Altman2009} we need an appropriate finite size scaling scheme.
The SDRG flow takes place in the space of the two parameters $\a$ and $u$ associated with the renormalized distributions of Josephson couplings and charging energies respectively. 
Only one of those parameters, $\alpha$, can be extracted from the numerical data.  
But, as we show below, this gives us sufficient information to formulate a finite size scaling scheme.

The essential idea is similar to a finite size scaling  procedure developed for the BKT critical point, where the RG operates in the space of stiffness and vortex fugacity, with only the stiffness being a directly measurable quantity.  As a preparatory step we solve the scaling equations given in Ref.~\onlinecite{AKPR} to obtain flow lines $u(\alpha,c)$, plotted in \fref{fig1}. Here $c$ is an integration constant which labels the flow lines. 
%
Next we bring the scaling equations to integral form
\begin{equation}
-\int_{\alpha_2}^{\alpha_1}\frac 1 {u(\alpha,c)} d\alpha-\log\left( \frac {\alpha_1+1}{\alpha_2+1}\right)+\log\left(\frac {L_2}{L_1}\right)=0
\label{eq:flow-eq-int-form}
\end{equation}
where $\alpha_1$ ($\alpha_2$) is the measured value at length $L_1$ ($L_2$).  There is then only one unknown in the above equation: the parameter $c$ which labels the flow lines. 

\begin{figure}[t]
\begin{center}
\includegraphics[width=\columnwidth]{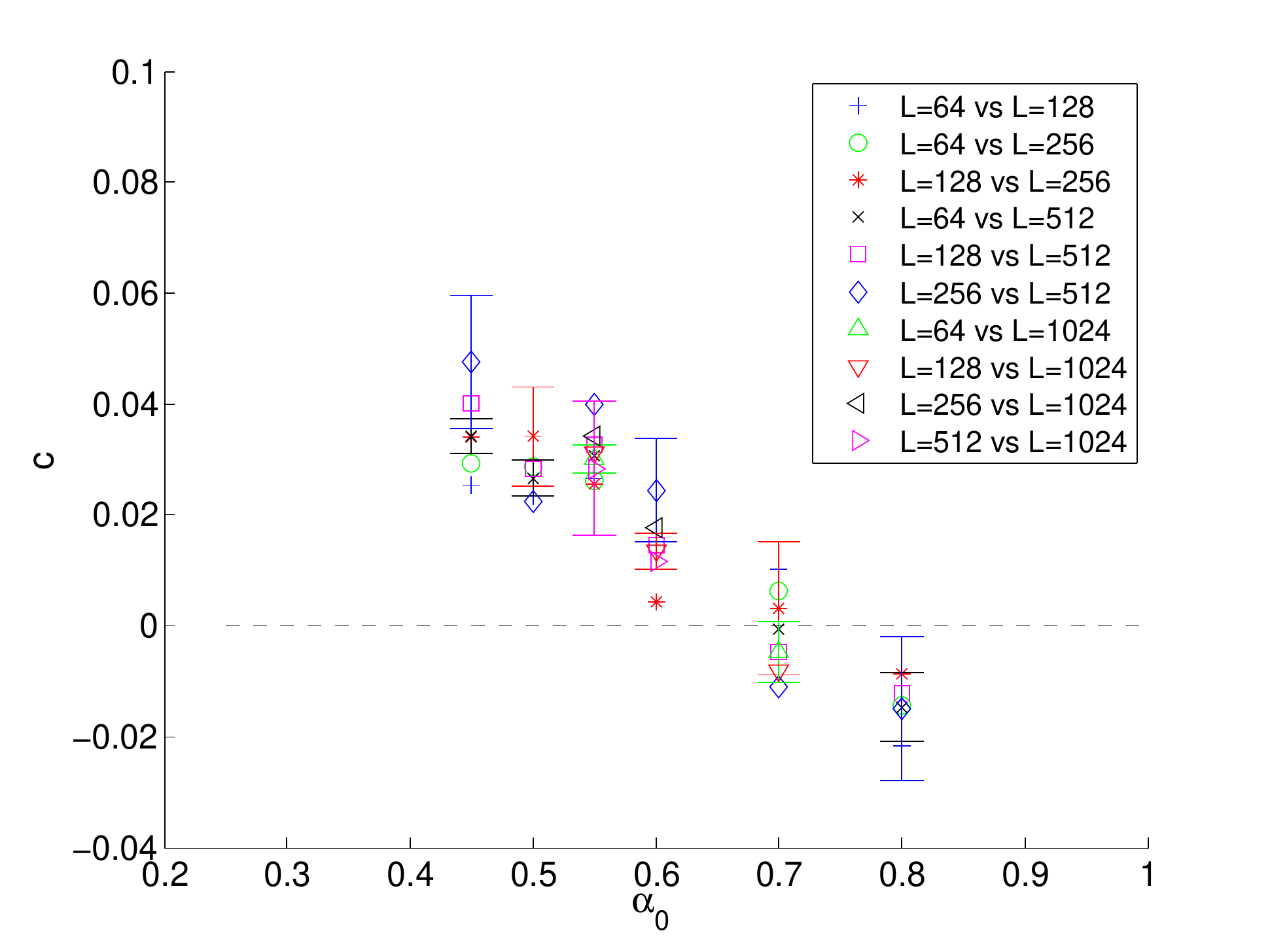}
\end{center}
\caption{
Solutions of integrated flow equations show good agreement with the strong-disorder fixed point. Solutions c are plotted for different combination of system sizes as a function of the tuning parameter $\alpha_0$. They agree within error bars. Here only the largest and smallest error bars for each $\alpha_0$ are shown. 
}
\label{fig3}
\end{figure}

Our finite-size scaling proceeds as follows: for each system size $L_i$ we measure $\alpha_i$. For each pair of system sizes $L_i$ and $L_j$ 
we solve \eqnref{eq:flow-eq-int-form} numerically for the parameter $c$. 
This tells us on which flow line, and thus in which phase, the system is: $c>0$ corresponds to the insulating phase, and $c<0$ to the superfluid phase. 
If the physics is indeed controlled by the strong disorder fixed point, then 
the solutions for different pairs of systems sizes (for the same bare system parameters) should match within error bars. \fref{fig3} shows that this is indeed the case. 
%
The error bars are created by shifting each value for $\alpha$ within its error bars, and recalculating $c$. To keep the figure clear, we only show the largest and the smallest error bar for each $\alpha_0$.
 
We observe good agreement between the different solutions for c, and conclude that our system is well described by the strong-disorder fixed point. In contrast, finite size scaling according to the standard BKT theory ~\cite{WeberHans1988} fails to converge for this system, see Appendix~\ref{BKT-scling}. 



\section{Conclusions}
We have provided direct numerical evidence for the strong-randomness fixed point of 
bosons in one dimension. This conclusion relied on using a new finite size scaling scheme that allowed direct comparison of the numerical data to the universal predictions of the strong disorder RG analysis [\onlinecite{AKPR}].
The alternative hypothesis of standard BKT scaling fails to describe the data. Our finite-size scaling is quite general and after a slight modification it works also for models without particle-hole symmetry, like the one studied in Ref.~\onlinecite{Pollet2013a} (see Appendix~\ref{appendix:scaling}). 
These results strongly suggest that the superfluid-to-insulator quantum phase transition changes universality class as the disorder strength is increased: from a BKT-like transition in the case of weak disorder, where the Luttinger parameter becomes universal at the critical point, to a different kind of phase transition, where the distribution of the superfluid stiffness becomes universal. 

It is interesting to note that there is a corresponding transition inside the superfluid phase from a clean superfluid fixed point described by a Luttinger liquid to a disordered superfluid characterized by broad power-law tails in the distribution of $1/\rho$. This behavior of $1/\rho$ leads to the anomalous localization properties discussed in 
Ref.~\onlinecite{Gurarie2008}. 

In this study we have concentrated on quantities for which the SDRG gives universal predictions. Hence we did not directly address the question of what is the value of the Luttinger parameter $K$ at the critical point. So far we do not have a reliable finite size scaling theory, which allows to extrapolate this value to the thermodynamic limit. Ref. \onlinecite{Pollet2013a} advocated an extrapolation formula based on completely classical field renormalization. However as we show in appendix \ref{appendix:classical} this formula significantly underestimates the value of the Luttinger parameter. 
The development of an improved extrapolation formula, which takes into account the flow of $\a$, i.e. quantum effects, is left for future work.

The bare disorder distributions used in the numerical simulations were close to the strong-disorder fixed point. 
It remains to be investigated under what circumstances an arbitrary disorder distribution flows to this strong-randomness critical point. A related open question is how the critical behavior changes in the intermediate regime between the weak disorder and strong disorder limits. An intriguing possibility is that this change is controlled by a new unstable fixed point separating the two regimes.



%

\noindent {\em Acknowledgments--}
We thank Lode Pollet, Boris Svistunov, Nikolay Prokof'ev, Daniel Podolsky, Lars Bonnes, Gil Refael, Anatoli Polkovnikov, and Yariv Kafri for illuminating discussions.
This work was supported by the ISF, the
Minerva foundation, the ERC UQUAM program, the NSF Grant No. PHY11-
25915 during a visit to the KITP-UCSB. 
E.A thanks the Aspen Center for Physics an  NSF Grant $\# 1066293$ for hospitality during the writing of this paper. E. A. acknowledges the hospitality of the Miller
Institute of Basic Research in Science.
S.P. acknowledges support from the Minerva Stiftung. 
The numerical simulations were carried out on the Weizmann ATLAS Grid cluster and the Weizmann WEXAC cluster. 

\begin{appendix}

\section{Double-peak structure of the distribution of superfluid stiffness}
\begin{figure}[tbh]
\begin{center}
\includegraphics[width=\columnwidth]{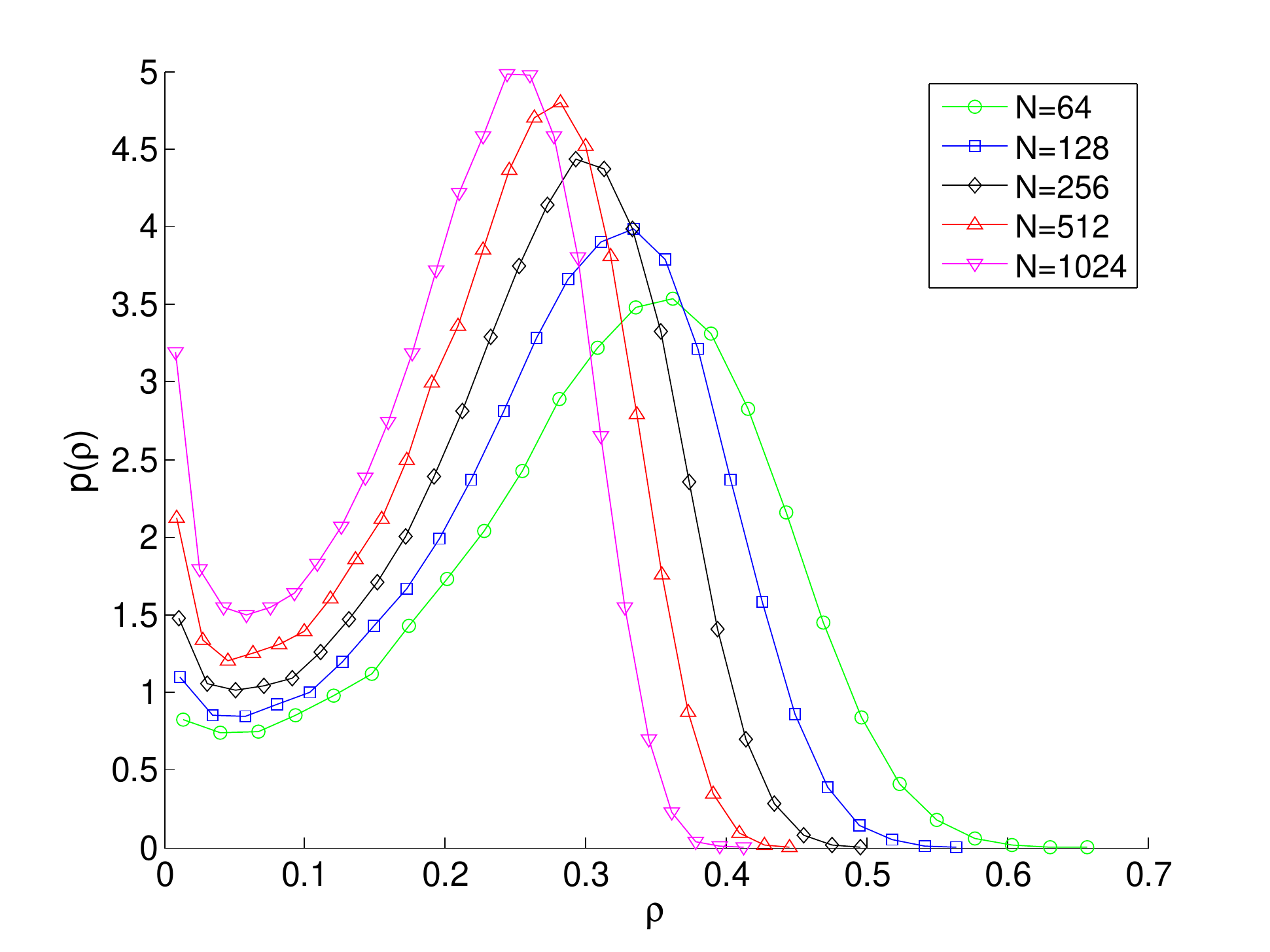}
\end{center}
\caption{
Distribution of superfluid stiffness obtained by carrying out the SDRG numerically. Here we have used $d=200000$ disorder realizations. The initial disorder distribution were the self similar fixed-point distributions with $u_0=0.2$ and $\alpha_0=0.4$. 
}
\label{figA1}
\end{figure}

The superfluid stiffness in Fig. 2a has a double-peak structure. This is indeed what we expect. We have calculated the superfluid stiffness expected from SDRG by carrying out the RG steps numerically, see \fref{figA1} for an example. Here also the double-peak structure is clearly visible.  Note that the SDRG operates on the Josephson array model given by Eq. (1). This is  different  than the current loop model on which the Monte-Carlo calculations are performed. Therefore we do not expect the distributions to match exactly. However the qualitative features  of the stiffness distribution are clearly similar, including the power-law tail and the double peak structure.

\section{Berezinskii-Kosterlitz-Thouless scaling fails}
\label{BKT-scling}
\begin{figure*}[t!]
\begin{center}
\includegraphics[width=\columnwidth]{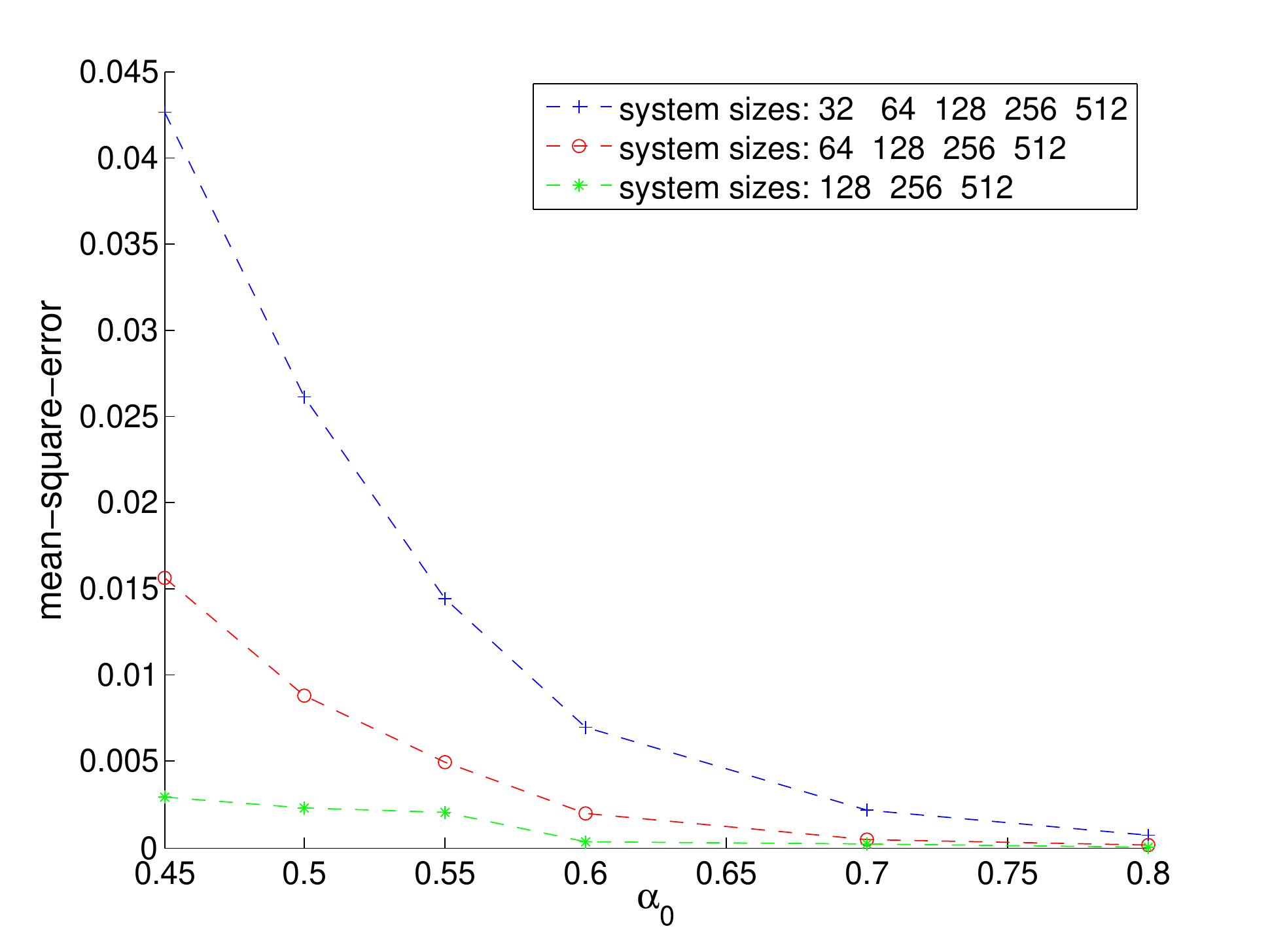}
\includegraphics[width=\columnwidth]{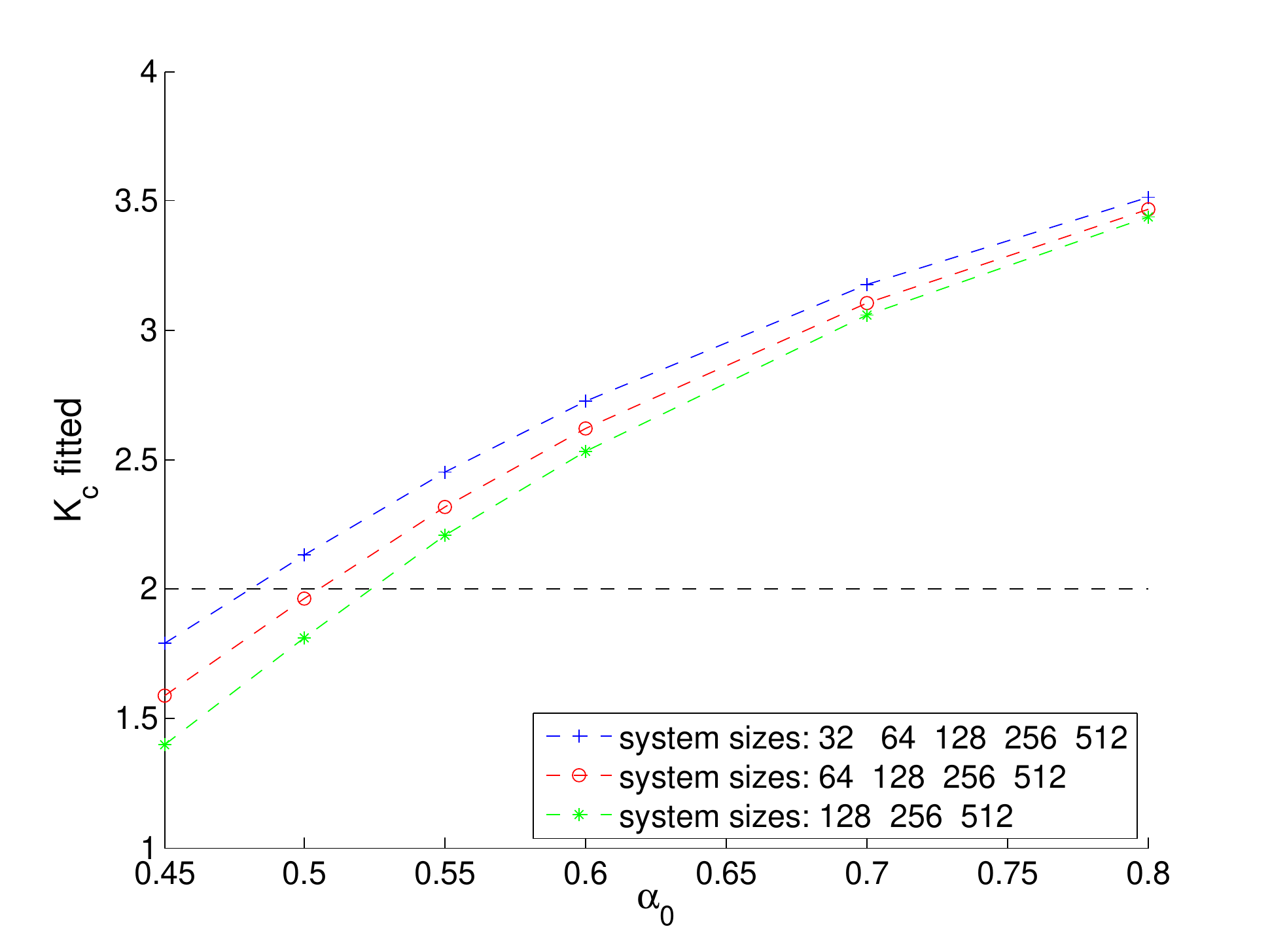}
\end{center}
\caption{
A standard BKT-scaling procedure, the Weber-Minnhagen fit, fails to locate a BKT phase transition. a) mean-square-error of the fit to \eqnref{eq:WM} should have a minimum at the transition, but here there is no minimum. b) fitted value for $K_c$; if this transition were in the BKT universality class, then we would find $K_c=2$ at the position of the minimum. 
Even if there might be a minimum for lager $\alpha_0$, this would be at a value of the Luttinger parameter $K>2$, and so this rules out the weak-disorder scenario. For this analysis the median of the Luttinger parameter $K$ was used. Similar results are obtained when using the average. 
}
\label{figA2}
\end{figure*}

Here we show that the transition of our model is not described by BKT scaling. Near the critical point of a BKT transition, the Luttinger parameter $K$ flows as a function of system size $L$ as
\beq
K(L)=K_c+\frac 1 {\ln(L)+C}
\label{eq:WM}
\eeq
where $K_c$ is the critical value of the Luttinger parameter at the transition, and $C$ is a non-universal constant. Following Ref.~\onlinecite{WeberHans1988}  we treat both $K_c$ and $C$ as fitting parameters, and fit our measured values of Luttinger parameter $K=\pi\sqrt{\rho_x \rho_{\tau}}$ 
to \eqnref{eq:WM}. 
If the transition were in the BKT universality class, then the mean-square error of the fit would have a clear minimum at the phase transition, and the corresponding fitted value of $K_c$ would be equals $2$ at the transition. 
As shown in \fref{figA2} this is not the case. 

\section{Scaling equations for the strong-disorder fixed point}
\label{appendix:scaling}
The scaling equations for the parameters $\alpha$ and $u$ from SDRG are~\cite{AKPR, Altman2009} (note that our parameter $u$ corresponds to the parameter $f_0$ in Refs.~\onlinecite{AKPR, Altman2009}). 
\begin{subequations}
\bea
\diff{u}{\Gamma}&=&-\alpha u - (\alpha+1) u^2\\
\diff{\alpha}{\Gamma}&=&-b(\alpha+1) u
\eea
\end{subequations}
Here $\Omega=\Omega_0e^{-\Gamma}$ is the energy cutoff. 
The value of the constant $b$ is the only difference between two different disordered models: 
 $b=1$ in the particle-hole symmetric case of a commensurate lattice potential at integer filling, and $b=1/2$ in case of chemical-potential disorder. 
These scaling equations are solved to obtain the flow lines $u(\alpha,b,c)$
\begin{equation*}
u(\alpha,b, c)=\frac 1 b e^{(\alpha+1)/b} \int_{(\alpha+1)/b}^{1/b}\frac{e^{-t}}{t} dt + e^{\alpha/b}-1+ce^{(\alpha+1)/b}.
\end{equation*}
Energy cutoff and renormalized length of the system are related as
\bea
\diff{l_\Gamma}{\Gamma}=l_ \Gamma(\alpha+1+u)
\eea
where $l_\Gamma$ is the average number of original sites per remaining cluster. In integral form these flow equations become
\begin{equation*}
-\int_{\alpha_2}^{\alpha_1}\frac 1 {u(\alpha,b,c)} d\alpha-\log\left( \frac {\alpha_1+1}{\alpha_2+1}\right)+b\log\left(\frac {L_2}{L_1}\right)=0.
\end{equation*}
For our model we have $b=1$; with $b=1/2$ our finite-size-scaling method also works for models with particle-hole symmetry, as the one studied in Ref.~\onlinecite{Pollet2013a}. We solve these flow equations numerically for the only unknown, $c$, see main text. 

\begin{figure}[h!]
\begin{center}
\includegraphics[width=\columnwidth]{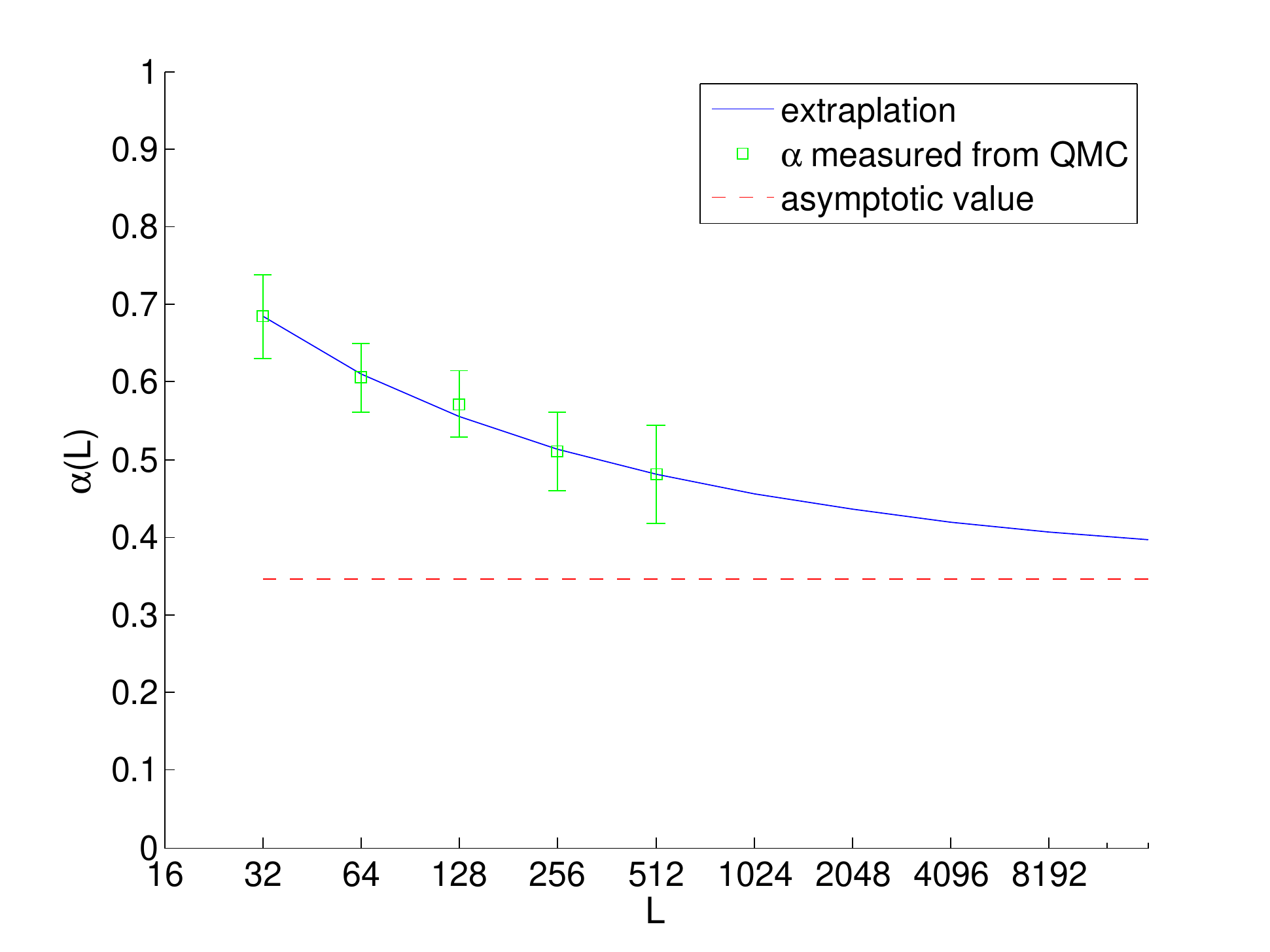}
\end{center}
\caption{
Monte Carlo data and extrapolation for $\alpha(L)$, for a system with bare parameter $\alpha_0=0.8$. The green squares are Monte Carlo results, the blue line shows the extrapolation from flow equations, and the red dashed line shows the asymptotic value $\lim_{L\rightarrow\infty} \alpha(L)$. 
}
\label{fig:extrapolate-alpha}
\end{figure}

Having determined $c$ we can use the flow equations to extrapolate the power $\alpha$ to infinite system sizes, $\alpha_{\infty}=\lim_{L\rightarrow\infty} \alpha(L)$, see \fref{fig:extrapolate-alpha}.

\section{How reliable is classical field renormalization?}\label{appendix:classical}

In order to obtain the value of the Luttinger parameter at the critical point from numerical simulations one needs to rely on a finite size scaling theory to extrapolate from finite size results to the thermodynamic limit.  In Ref.~\onlinecite{Pollet2013a}, Pollet \etal~have proposed an extrapolation formula based on the scaling of the stiffness in a classical Josephson junction array. Using such  extrapolation, these authors conclude that the Luttinger parameter evaluated at the transition from their numerical results is not above the universal value predicted by the weak disorder theory\cite{Giamarchi1987}. In this appendix we assess how reliable is an extrapolation based on the assumption of classical field renormalization. 

Classical field renormalization, as applied in Ref. ~\onlinecite{Pollet2013a} describes a classical Josephson junction array with random couplings, which are distributed according to a power-law 
$p(J)\propto J^{\alpha}$ for $J<\Omega_0$, and zero otherwise (this is the same as the fixed-point distribution from SDRG; note that our $\alpha$ corresponds to $-\zeta$ in Ref.~\onlinecite{Pollet2013a}). 
%

The median of the inverse stiffness $\rho^{-1}_x(L)=\sum_{i=1}^{L}\frac 1 {J_i}$ in this classical model scales as ~\cite{Pollet2013a}
\be
\rho_x^{-1}(L)=\rho_{\rm cl}^{-1}-\frac B {\alpha} L^{-\alpha_{cl}}
\label{eq:class-flow}
\ee
where $B$ and $\rho_{\rm cl}^{-1}$ are constants, which are used as additional fitting parameters. Hence there are 3 fitting parameters used to fit a set of points $\{\rho(L_1),\ldots,\rho(L_N)\}$ generated from simulations of a model with a fixed distribution of (bare) coupling constants at different system sizes. A fit to the above equation allows to identify a power $\alpha_{cl}$ of the distribution, and to extrapolate the superfluid stiffness to an infinite system size, and thus to extrapolate the Luttinger parameter $K=\pi \sqrt{\rho_x \rho_{\tau}}$ (note that the compressibility $\kappa=\rho_{\tau}$ barely changes as a function of system size in the superfluid phase).
\begin{figure}[tbh]
\begin{center}
\includegraphics[width=\columnwidth]{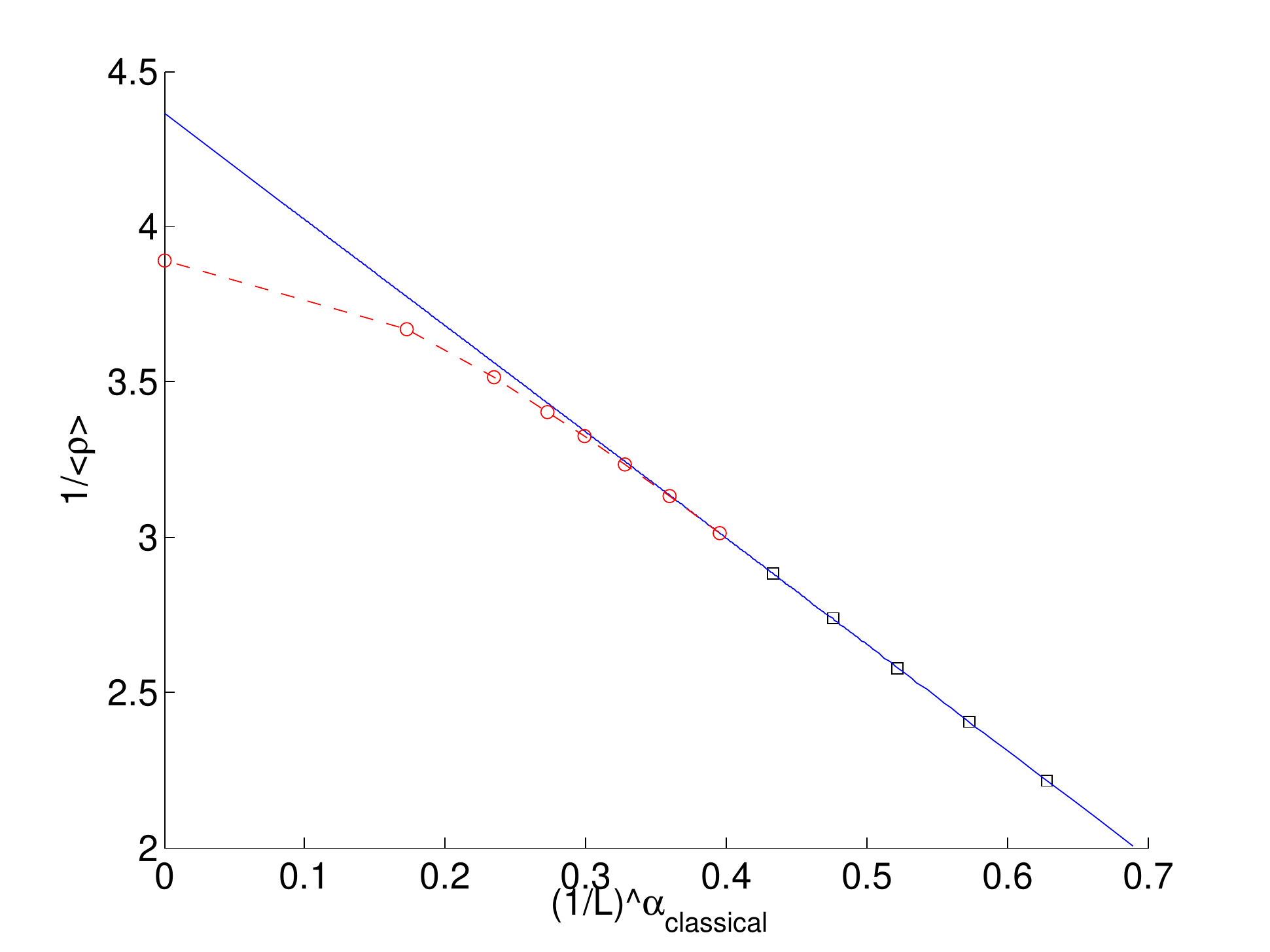}
\end{center}
\caption{
A simple toy model shows that the classical field renormalization flow cannot be used reliably when the exponent $\alpha$ changes as a function of system size, see text. 
We used the values of $\alpha$ obtained from QMC simulations in the superfluid phase, and extrapolated to larger system sizes.
Each system size has its own parameter $\alpha(L)$.  
The blue squares and red circles correspond to the inverse superfluid stiffness computed for the classical Josephson junction array model.  
The blue line shows the extrapolation from classical flow for system sizes up to $L=512$ (blue squares). The classical field extrapolation seems to work well for these small systems, but it underestimates the value of the superfluid stiffness for an infinite system size. Note that the $x$ axis has been rescaled using the parameter $\alpha_{\rm cl}$ obtained from the classical flow. 
}
\label{figA4}
\end{figure}

As we argue below, this extrapolation for $K$ is not reliable if the power $\alpha$ characterizing the distribution of $J$ changes as a function of system size. 
We demonstrate this with a simple toy model.
For each given system size, we take a classical Josephson junction array model with a power-law distribution of $J$, $P(J)\propto J^{\a(L)}$ for $J<\W_0$, but where each system size has its own exponent $\alpha(L)$. If we are given the scaling law $\a(L)$, then in this classical model we can easily compute $\rho(L)$ numerically for given system sizes. We also know the exact superfluid stiffness of the infinite system to be $\rho_\infty=\W_0\alpha(\infty)/(\a(\infty)+1)$. At the same time, given the numerical results for $\rho(L)$ from several system sizes (e.g. $L=32,64,128,256,512$) we can attempt to fit this data assuming the classical scaling formula (\ref{eq:class-flow}) with a single exponent $\a_{\rm cl}$ in spite of the data being generated from models with $L$ dependent $\a$. Most importantly, from the fit we can obtain an estimate of the extrapolated stiffness in the thermodynamic limit $\rho_{\rm cl}$ which we can compare against the exact known value.

Let us carry out this procedure using the values $\alpha(L)$ obtained in the MC calculations conducted in the superfluid phase (we use the simulations with the bare parameter $\alpha_0=0.8$). We can readily extrapolate these values to any system size $L$, all the way to the thermodynamic limit, using the scaling $\a(L)$ predicted by the SDRG. The comparison between the predicted $\a(L)$ scaling and the MC result, shown in Fig. \ref{fig:extrapolate-alpha}, requires no other fitting parameter except the number $c$ which parameterizes the RG trajectory the system is flowing on. This parameter  was already extracted from the data in the main text.

In the toy model we computed the median of the inverse stiffness for each system size numerically from 50000 realizations of the classical Josephson array . We then fit the results obtained from the smaller systems sizes (32, 64, 128, 256, 512) to \eqnref{eq:class-flow} and obtain a good fit with fitting parameters $\alpha_{\rm cl}=0.13$ and 
$\rho_{\rm cl}=0.23$, while the correct asymptotic values in this model are $\alpha(\infty)=0.35$ and $\rho_{\infty}=0.26$. 
More generally, we find that the fit to the classical flow systematically underestimates the power $\alpha$, as well as the value of $\rho_{\infty}$ extrapolated to infinite system sizes, see \fref{figA4}. 

We thus conclude that when the power $\alpha$ changes as a function of system size, as it does near the transition and in the insulator, then the true value for the Luttinger parameter is larger (in our the example shown, larger by 50\% ) than the Luttinger parameter extrapolated using the classical flow assumption. Hence, the latter  is not a reliable extrapolation for the true Luttinger parameter at infinite system size. 

\end{appendix}


\end{document}